\documentstyle[12pt]{article}
\newcommand{\pp}{{=\!\!\!|}}
\begin{document}
\newcommand{\eqn}[1]{eq.(\ref{#1})}
\renewcommand{\section}[1]{\addtocounter{section}{1}
\vspace{5mm} \par \noindent
  {\bf \thesection . #1}\setcounter{subsection}{0}
  \par
   \vspace{2mm} } 
\newcommand{\sectionsub}[1]{\addtocounter{section}{1}
\vspace{5mm} \par \noindent
  {\bf \thesection . #1}\setcounter{subsection}{0}\par}
\renewcommand{\subsection}[1]{\addtocounter{subsection}{1}
\vspace{2.5mm}\par\noindent {\em \thesubsection . #1}\par
 \vspace{0.5mm} }
\renewcommand{\thebibliography}[1]{ {\vspace{5mm}\par \noindent{\bf
References}\par \vspace{2mm}}
\list
 {\arabic{enumi}.}{\settowidth\labelwidth{[#1]}\leftmargin\labelwidth
 \advance\leftmargin\labelsep\addtolength{\topsep}{-4em}
 \usecounter{enumi}}
 \def\newblock{\hskip .11em plus .33em minus .07em}
 \sloppy\clubpenalty4000\widowpenalty4000
 \sfcode`\.=1000\relax \setlength{\itemsep}{-0.4em} }
{\hfill{BRX-TH-419, VUB/TENA/97/4, hep-th/9706218}}
\vspace{1cm}
\begin{center}
{\bf THE QUANTUM GEOMETRY OF $N=(2,2)$ NON-LINEAR $\sigma$-MODELS }

\vspace{1.4cm}
M.T. GRISARU${}^*$, M. MASSAR${}^\dagger$,
A. SEVRIN${}^\dagger$ and J. TROOST${}^\dagger$\footnote{Aspirant NFWO} \\
${}^*${\em Physics Department, Brandeis University,} \\
{\em Waltham, MA, USA} \\
${}^\dagger${\em Theoretische Natuurkunde, Vrije Universiteit Brussel} \\
{\em Pleinlaan 2, B-1050 Brussel, Belgium} \\
\end{center}
\centerline{ABSTRACT}

\vspace{- 4 mm}  

\begin{quote}\small
We consider a general $N=(2,2)$ non-linear $\sigma$-model in $(2,2)$
superspace. Depending on the details of the complex structures involved,
an off-shell description can be given in terms of chiral, twisted chiral
and semi-chiral superfields. Using superspace techniques, we derive the
conditions the potential has to satisfy in order to be
ultra-violet finite at one loop. We pay particular
attention to the effects due to the presence of
semi-chiral superfields. A complete description of $N=(2,2)$ strings follows
from this.
\end{quote}
\baselineskip18pt
\addtocounter{section}{1}
\noindent
Ever since Zumino's observation \cite{bruno}  that the target manifold of a
four-dimensional supersymmetric $\sigma$-model needs to be K\"ahler, it
has been recognized that supersymmetry and complex geometry go hand in
hand. In two dimensions this relation becomes particularly rich. In this paper,
we study
non-linear $\sigma$-models with $N=(2,2)$ supersymmetry. Such models are not
only the
building blocks for type $II$ superstrings, but they describe
the matter sector of $N=(2,2)$ string theories as well. Classically, $(2,2)$
supersymmetry requires the existence of two covariantly constant complex
structures, which are such
that the metric is hermitean for both. The simplest choice, both complex
structures equal,
implies that the target manifold is K\"ahler. The other cases generalize
the notion of K\"ahler geometry in the sense that the geometry can always
locally be described in terms of a single potential.
For such models the ultra-violet divergences are much milder than what can
be expected from power counting. However, requiring ultra-violet
finiteness imposes further restrictions on this potential. Up till now,
this has only been studied for target manifolds where
the supersymmetry algebra closes off-shell in $(1,1)$ superspace. This is
equivalent to the requirement that both complex structures commute.
In the present paper, we examine the generic case, where off-shell closure
requires a $(2,2)$ superspace formulation.

An arbitrary $N=(1,1)$ supersymmetric non-linear $\sigma$-model has an
action in $(1,1)$ superspace \cite{GHR} given by\footnote{
We take ${\tilde D}_+\equiv\frac{\partial}{\partial{\tilde \theta}^+}+
{\tilde \theta}^+\partial_\pp$ and
${\tilde D}_- \equiv \frac{\partial}{\partial {\tilde \theta}^-}+
{\tilde \theta}^- \partial_=$. Our
two-dimensional metric is such that its only nonvanishing component is
$g_{\pp\, =} =1$.}
\begin{eqnarray}
{\cal S}=\frac{1}{2\pi}\int d^2 x d^2 {\tilde \theta}
\left(g_{ab}+b_{ab}\right){\tilde D}_+\phi^a {\tilde D}_-
\phi^b,\label{staction}
\end{eqnarray}
where the metric on the target manifold is $g_{ab}$ while $b_{ab}=-b_{ba}$ is
the torsion potential:
\begin{eqnarray}
T^a{}_{bc}\equiv-\frac 3 2 g^{ad}b_{[bc,d]}.\label{Sdef}
\end{eqnarray}
This model is $N=(2,2)$ supersymmetric, provided there exists two complex
structures $J_+$
and $J_-$, which are
covariantly constant,
\begin{eqnarray}
\nabla_c^+J_+^a{}_b=\nabla_c^- J_-^a{}_b=0, \label{herm}
\end{eqnarray}
where $\nabla^\pm$ denotes
covariant differentiation using the  $\Gamma_\pm$ connections:
\begin{eqnarray}
\Gamma^a_{\pm bc}\equiv  \left\{ {}^{\, a}_{bc} \right\} \pm
T^a{}_{bc},
\end{eqnarray}
the first term being the standard Levi-Cevita connection.
Finally, the metric should be hermitean w.r.t. both complex structures.
For a four-dimensional target manifold this simplifies as one can show
that any covariantly constant {\it almost} complex structure for
which the metric is hermitean is automatically a complex structure
\cite{old}, i.e. the Nijenhuis tensor vanishes trivially.
On-shell, one
gets the standard $N=(2,2)$ supersymmetry algebra, while the off-shell
non-closure
terms are proportional to the commutator $[J,\bar J]$.

Building on the results of \cite{martinnew},
it was argued in \cite{icproc} that, in order to achieve a manifest
$(2,2)$ supersymmetric description of these models, {\it i.e.} a
formulation in $(2,2)$ superspace, chiral, twisted chiral \cite{GHR} and
semi-chiral
\cite{BLR}
fields are sufficient. More explicitly:
the tangent space at any point of the target manifold
can be decomposed into three subspaces: $\ker (J_+ -
J_-)\oplus \ker (J_++J_-) \oplus (\ker[J_+,J_-])^\perp $. These subspaces are
conjectured to be integrable to chiral, twisted chiral and semi-chiral
coordinates resp. In $N=(2,2)$ superspace, we have four
fermionic coordinates, denoted by
$\theta^+$, $\bar\theta^+$, $\theta^-$ and $\bar\theta^-$ with covariant
derivatives $D_+$, $\bar D_+$, $D_-$ and $\bar D_-$. The only
non-vanishing anticommutators are: $\{ D_+,\bar D_+\}=\partial_\pp$ and
$\{ D_-,\bar D_-\}=\partial_=$. Note that the $N=(1,1)$ fermionic
coordinates previously introduced are given by the real parts of the
$N=(2,2)$ coordinates, $\tilde{\theta}^\pm=(\theta^\pm+\bar\theta^\pm)/2$.
The three types of superfields are defined through the following
constraints.

\vspace{.3cm}

\noindent{\it i.} \underline{Chiral superfields}: $z^a$, $\bar z^{\bar a}$;
$a,\,\bar a\in
\{1,\cdots d_c\}$.\\
$D_\pm z^a=\bar D_\pm \bar z^{\bar a} =0$.\\
{\it ii.} \underline{Twisted chiral superfields}: $w^m$, $\bar w^{\bar m}$;
$m,\,\bar m\in
\{1,\cdots d_t\}$\\
$D_+ w^m= \bar D_- w^m=\bar D_+\bar w^{\bar m}=D_- \bar w^{\bar m} =0$.\\
{\it iii.} \underline{Semi-chiral superfields}\footnote{Our notation is
slightly misleading. While $J_+$ and $J_-$ are both diagonal for chiral
and twisted chiral superfields, this is not so for semi-chiral
superfields. This is evident from the fact that semi-chiral fields
parametrize $(\ker[J_+,J_-])^\perp $.}: $ r^\alpha $, $\bar r^{\bar\alpha }$,
$ s^{\tilde\alpha} $, $\bar s^{\tilde{\bar\alpha} }$;
$\alpha ,\, \bar\alpha ,\, \tilde\alpha ,\, \tilde{\bar\alpha}\in\{
1,\cdots , d_s\}$.\\
$D_+  r^\alpha=\bar D_+  \bar r^{\bar\alpha }=\bar D_-  s^{\tilde\alpha} =
D_-  \bar s^{\tilde{\bar\alpha} }=0$.

\vspace{.3cm}

\noindent The constraints reduce the number of components of chiral and twisted
chiral superfields to those of an $N=(1,1)$ superfield. Semi-chiral
superfields have twice as many components, half of which
are auxiliary. This is not so surprising as chiral and twisted chiral
superfields parametrize $ \ker[J_+,J_-]=
\ker (J_+ - J_-)\oplus \ker (J_++J_-) $ where the $N=(2,2)$ algebra
closes off-shell.

The action in $(2,2)$ superspace is
\begin{eqnarray}
{\cal S}=\int d^2x d^4\theta K(z,\bar z, w, \bar w,
 r,\bar r,  s,\bar s), \label{n2ac}
\end{eqnarray}
with $K$ a real potential. The potential is determined modulo a generalized
K\"ahler
transformation:
\begin{eqnarray}
K\simeq K+ f(z,w, r)+\bar f(\bar z, \bar w,\bar  r)+ g(z,\bar w,
\bar s)+\bar g(\bar z, w,  s).\label{gkt}
\end{eqnarray}
Starting from eq. (\ref{n2ac}), one can pass to $N=(1,1)$ superspace. Upon
elimination of the
auxiliary fields, one gets, by comparing the result to eq. (\ref{staction})
and the supersymmetry transformation rules,
explicit expressions for the metric,
torsion and complex structures in terms of the potential \cite{icproc}.
Various explicit examples are known.
K\"ahler manifolds are described using chiral fields only. The
$SU(2)\times U(1)$ Wess-Zumino-Witten (WZW) model can be described either in
terms of a
chiral and a chiral multiplet \cite{RSS} or in terms of one semi-chiral
multiplet \cite{icproc,martinnew}. This ambiguity reflects the freedom one has
in choosing
the left and
right complex structures. The WZW model on $SU(2)\times SU(2)$ is
described in terms of a semi-chiral and a twisted chiral field \cite{icproc}.
Even hyper-K\"ahler manifolds can be described in terms of semi-chiral
coordinates \cite{icproc}. Indeed, choosing $J_+$ and $J_-$ such that
$\{J_+,J_-\}=0$, one gets $\ker [J_+, J_-]=\emptyset$.

Conformal invariance of these models puts strong
restrictions on the allowed potentials $K$ in eq. (\ref{n2ac}).
The requirement that the one loop $\beta$-function vanishes \cite{Rdil},
\begin{eqnarray}
R^+_{ab}+2\nabla^-_a\partial_b\Phi=0 , \label{bet}
\end{eqnarray}
with $\Phi$ the dilaton, yields various dilaton configurations together
with differential equations for the potential\footnote{Our
convention for the curvature tensor is
$R^a_{\ bcd}\,=\,\Gamma^a_{\ bd,c}\,+\,\Gamma^a_{\ ec}\Gamma^e_{\ bd}
\,-\,(c\,\leftrightarrow\,d)$. Pending upon whether we use the $\Gamma_+$
or the $\Gamma_-$ connection, we get two curvature tensors related by
$R^+_{abcd}\,=\,R^-_{cdab}$. The
Ricci tensors $R^\pm_{ab}$ are defined by
$R^\pm_{ab} \equiv R^{\pm}{}^c{}_{acb}$
and $R^+_{ab}=R^-_{ba}$. Covariant derivatives are taken as
$\nabla^+_bV^a\equiv
V^a{}_{,b}+\Gamma^{+a}{}_{cb}V^c$ and $\nabla^+_bV_a\equiv
V^a{}_{,b}-\Gamma^{+c}{}_{ab}V^c$. }.

So our task looks simple: starting from eq. (\ref{n2ac}) we pass to $N=(1,1)$
superspace, eliminate the auxiliary fields, and compare with eq.
(\ref{staction}).
In that way, we get explicit expressions for metric and torsion in terms of the
potential,
to use in
the $\beta$-function eq. (\ref{bet}) which can then be analysed.
This is indeed straightforward as long as only
chiral and twisted chiral fields are present.
Having only chiral fields and choosing the dilaton constant, yields
\cite{muk}
\begin{eqnarray}
\det K_{a\bar b}=1,  \label{aa}
\end{eqnarray}
where $K_{a\bar b} $ stands for $\partial^2 K/\partial z^a\partial z^{\bar
b}$.
When both chiral and twisted chiral superfields are present and the dilaton is
chosen as $\Phi=(1/2)\ln\det K_{a\bar b}$,
eq. (\ref{bet}) is equivalent to \cite{kiritsis}
\begin{eqnarray}
\det K_{a\bar b}=\det(- K_{m \bar n}),     \label{bb}
\end{eqnarray}
More general solutions, where the dilaton is a potential
for a holomorphic Killing vector, were studied in \cite{elias}.
Once semi-chiral fields enter the game, the metric and
torsion (see e.g. \cite{icproc} or \cite{BLR}) become so complicated that
the whole program becomes technically unfeasable.
The way out of course, is to recompute the $\beta$-functions, but now
directly in $(2,2)$ superspace. At present, there is no good understanding
of the dilaton in the presence of semi-chiral sperfields, so in this paper
we limit ourselves to  the study of a necessary condition for conformal
invariance: ultra-violet finiteness. In fact, ultra-violet finiteness for
chiral fields is precisely equivalent to eq. (\ref{aa}), while for chiral
and twisted chiral fields it is equal to eq. (\ref{bb}).

Some care is required in setting up the Feynman rules. As an example, we
compute the one loop UV divergence for a potential which depends on one
semi-chiral multiplet.
The only subtle part is the construction of the free field propagators.
A particularly convenient
potential is given by $K_0=1/2( r\bar r+ s\bar s)+
 r\bar s + s \bar r$. We introduce sources, $j$, $\bar j$, $k$ and
$\bar k$, which are unconstrained superfields. The complete action can be
rewritten as
\begin{eqnarray}
{\cal S}&=&\int d^2x d^4\theta\left( K_0+j r+\bar j \bar r +k  s +\bar k\bar
 s \right)\nonumber\\
&=&\int d^2x d^4\theta\left(\frac 1 2 (  r\bar r+ s\bar s)  +
\frac{D_+\bar D_+}{\partial_\pp} r
\frac{D_- \bar D_-}{\partial_=} \bar s   + \frac{\bar D_- D_-}{\partial_=}
 s  \frac{\bar D_+ D_+}{\partial_\pp} \bar r+ \right.\nonumber\\
&&\left.  r \frac{\bar D_+ D_+}{\partial_\pp}j+
\bar r \frac{ D_+\bar D_+}{\partial_\pp}\bar j+
 s \frac{ D_-\bar D_-}{\partial_=}k+
\bar s \frac{ \bar D_- D_-}{\partial_=}\bar k\right)\nonumber\\
&=& \int d^2x d^4\theta\left( K_0(\hat{ r},\hat{\bar r},\hat{ s},
\hat{\bar s})+\frac
4 3 \bar j \frac{\bar D_+D_+D_-\bar D_-}{\partial_{\pp}\,\partial_=}k
+\frac 4 3 \bar k\frac{\bar D_+D_+D_-\bar
D_-}{\partial_{\pp}\,\partial_=}j+\right.\nonumber\\
&&\left. \bar j\frac{\bar D_+D_+}{\partial_{\pp}}(2-\frac 8 3
\frac{D_-\bar D_-}{\partial_=})j+
\bar k\frac{D_-\bar D_-}{\partial_{=}}(2-\frac 8 3
\frac{\bar D_{+}D_+}{\partial_{\pp}})k
\right).\label{props}
\end{eqnarray}
The slightly unconventional rewriting of $K_0$ in the second line which uses
the
semi-chiral properties, is
needed in order that the shifted fields in the standard Gaussian integral,
\begin{eqnarray}
\hat r&\equiv & r + \frac{D_+\bar D_+}{\partial_{\pp}}(2-\frac 8 3
\frac{\bar D_- D_-}{\partial_=})j+\frac 4 3 \frac{D_+\bar D_+\bar D_-D_-}
{\partial_{\pp}\,\partial_=}k,\nonumber\\
\hat s&\equiv&  s + \frac{\bar D_-D_-}{\partial_{=}}(2-\frac 8 3
\frac{D_+\bar D_+}{\partial_{\pp}})k+\frac 4 3 \frac{D_+\bar D_+ \bar D_-D_-}
{\partial_{\pp}\,\partial_=}j,
\end{eqnarray}
satisfy the same constraints as
$ r$ and $ s $.  From eq. (\ref{props}), we immediately obtain the
free field propagators.

Next, we consider a general potential $K( r',\bar r', s ',\bar s
')$. In order to perform the one-loop calculation, we make a linear
background-quantum splitting: $ r'= r_c+ r$ and $ s '= s _c+ s $, where
$ r_c$ and $ s _c$ are fixed background configurations and $ r$ and
$ s $ describe the quantum fluctuations around this background.
Expanding the potential around the background yields the interaction terms.
As usual, the terms linear in the quantum fields are irrelevant, while
for the one-loop computation, only the quadratic terms contribute:
\begin{eqnarray}
{\cal S}_{int}&=&\int d^2xd^4\theta \left( (K_{ r\bar r}-\frac 1 2)  r\bar r +
(K_{ s \bar s }-\frac 1 2) s \bar s +(K_{ r\bar s }-1)  r\bar s +
(K_{ s \bar r}-1)  s \bar r +\right.\nonumber\\
&&\left. \frac 1 2 K_{ r r}\, r r +
\frac 1 2 K_{\bar r\bar r}\,\bar r\bar r+  \frac 1 2 K_{ s s}\, s s+
\frac 1 2 K_{\bar s \bar s }\,\bar s \bar s  +\mbox{ 3rd order
terms}\right).\label{int}
\end{eqnarray}
The $D$-algebra is greatly simplified by the observation in \cite{m1}:
the background dependent terms can be treated as constants. Indeed, as the
superspace
measure is dimensionless, the counterterms are necessarily dimensionless
themselves.
Therefore, any derivative hitting the background terms cannot give rise to UV
divergent
contributions. A similar dimensional argument implies that the last four
interaction
terms in eq. (\ref{int}) do not contribute to the UV divergence as the $D$
algebra
always gives rise to derivatives acting on the background dependent terms.
Furthermore, one
argues along exactly the same lines as in \cite{m1}, that non-covariant
contributions to the counterterms will be absent.
The one-loop computation is now completely standard (using the techniques
described in
e.g. \cite{book}) and leads to the
counterterm:
\begin{eqnarray}
{\cal S}^{(1)}=\frac{ 1}{ 2\pi\varepsilon} \int d^2xd^4\theta
\ln\frac{1}{3} \frac{K_{ r\bar s }K_{ s \bar r}- K_{ r\bar r}K_{ s \bar s }}
{K_{ r\bar r}K_{ s \bar s }-K_{ r s }K_{\bar r\bar s }}.
\label{ct}
\end{eqnarray}
{}From this we get the condition for UV finiteness at one loop:
\begin{eqnarray}
\frac{K_{ r\bar s }K_{ s \bar r}- K_{ r\bar r}K_{ s \bar s }}
{K_{ r\bar r}K_{ s \bar s }-K_{ r s }K_{\bar r\bar s }} =
\pm |F( r)|^2 |G( s )|^2, \label{d4con}
\end{eqnarray}
where $F$ and $G$ are arbitrary functions of $ r$ and $ s $ resp.
Note that there is no coordinate transformation compatible with the
constraints, which can remove $|F( r)|^2 |G( s )|^2$.
The $+$ and $-$ signs hold for manifolds with $(4,0)$ and $(2,2)$
signature\footnote{
Though our computation was done for $(4,0)$ signature, it can easily be
redone for $(2,2)$ signature. One just replaces the quadratic potential by
$K_0=1/2( r\bar r- s\bar s)+
 r\bar s + s \bar r$ and one gets the same counterterm as in eq.
(\ref{ct}), except that the factor 3 in the denominator is replaced by a
factor $-5$.}.
This result can immediately be verified using some examples which are
known to be UV finite. The WZW model on $SU(2)\times U(1)$ is described by
a semi-chiral multiplet \cite{icproc,martinnew} with potential
\begin{eqnarray}
K=- r \bar{ r} + \bar{ r} \bar{ s } +  r  s
-2 i \int^{\bar{ s }- s } dx \, \ln ( 1+ \exp \frac{i}{2} x),
\end{eqnarray}
and we find $F=1$ and $G=\exp(-i s /2)$. Another class of interesting examples
are the
4-dimensional hyper-K\"ahler manifolds where $J_+$ and $J_-$ are choosen
to be anti-commuting. The potential satisfies then \cite{icproc}
$ |K_{ r s }|^2+|K_{ r\bar  s }|^2=2K_{ s \bar  s }K_{ r\bar r}$ and
$F=G=1$.

The computation can immediately be generalized to the case where $d_s$
semi-chiral
multiplets are present. The one-loop counterterm reads then
\begin{eqnarray}
{\cal S}^{(1)}=\frac{ 1}{2\pi \varepsilon} \int d^2xd^4\theta\,
\ln\frac{\det {\cal N}_2}{\det (\sqrt{-3} {\cal N}_1)} ,
\end{eqnarray}
where
\begin{eqnarray}
{\cal N}_1\equiv \left( \begin{array}{cc} K_{\widetilde\alpha \beta} &
K_{\widetilde\alpha \widetilde{ \bar\beta}}\\
K_{\bar\alpha \beta} & K_{\bar\alpha \widetilde{ \bar\beta}}\end{array}\right),
\qquad
{\cal N}_2\equiv \left( \begin{array}{cc} K_{\widetilde\alpha \bar\beta} &
K_{\widetilde\alpha  \widetilde{\bar\beta}}\\
K_{\alpha \bar\beta} & K_{\alpha
\widetilde{\bar\beta}}\end{array}\right).\label{ndef}
\end{eqnarray}
In \cite{icproc}, it was shown that non-degeneracy of the target manifold
metric is equivalent to $\det {\cal N}_1 \neq 0$ and  $\det {\cal N}_2 \neq 0$.

Finally, one can also repeat the calculation for a general potential, eq.
(\ref{n2ac}), which depends on all three types of superfields. One finds
the counterterm
\begin{eqnarray}
{\cal S}^{(1)}=\frac{ 1}{2\pi \varepsilon} \int d^2xd^4\theta\,
\ln\frac{\det (-K_{m\bar n})\det {\cal N}'_2}{\det K_{a\bar b}\det (\sqrt{-3}
{\cal N}'_1)} ,
\end{eqnarray}
where ${\cal N}'_1$ and ${\cal N}'_2$ are similar to ${\cal N}_1$ and ${\cal
N}_2$ in eq.
(\ref{ndef}) except that in ${\cal N}'_1$ one has to write the $d_s\times d_s$
matrices $(K_{AB}) -(K_{Aa})(K_{\bar a b})^{-1}(K_{\bar b B})$ instead of
the matrices $K_{AB}$, while in ${\cal N}'_2$ one has
$(K_{AB}) -(K_{Am})(K_{\bar m n})^{-1}(K_{\bar n B})$ instead of $K_{AB}$.
In these expressions,
the capital letters denote the indices appearing in eq. (\ref{ndef}),
while the small indices denote derivatives w.r.t. chiral or twisted chiral
fields, following the notation introduced in the definition of these
fields. Again, this result can be verified using a non-trivial example.
In \cite{icproc}, $SU(2)\times SU(2)$ was described in terms of a semi-chiral
multiplet $ r$, $ s$ and a chiral field $z$ with potential
\begin{eqnarray}
K&=&-z\bar z+z\bar r+\bar z r+i s z -i \bar s \bar z+i \bar s
\bar r-i s  r\nonumber\\
&&-i\int^{\bar r- r}dy\ln (1-\exp i y)-i
\int^{\bar s - s }dy\ln (1-\exp iy).
\end{eqnarray}
Using the previous expression and taking eq. (\ref{gkt}) into account,
one easily verifies that this
potential is indeed finite at one loop.

{}From the results in \cite{tor}, one expects new counterterms to appear from
four
loops on. Using arguments similar to those in \cite{nem} one can
presumably argue that a deformation of the potential can be defined order
by order so that they vanish.

Finally we consider the four dimensional case which is relevant for the
study of $N=(2,2)$ strings. There are three possible backgrounds: those given
in
terms of two chiral fields, those described by one chiral and one twisted
chiral and finally those consisting of a single semi-chiral multiplet.
Note that having two twisted chiral fields only is isomorphic to having
only two chiral fields.

In \cite{ov}, $N=(2,2)$ strings in a
K\"ahler background were studied\footnote{In \cite{ov} a different path was
taken. The
effective action was obtained by directly studying the
scattering amplitudes. The dynamics was given by a scalar field $\Phi$,
satisfying
the Plebanski equation. This can be
obtained from the Monge-Amp\`ere equation by setting $K=K_0+\Phi$ with
$K_0=z_1\bar z_1\pm z_2\bar z_2$.}. A description in terms of two chiral fields
can be given and ultra-violet finiteness requires that the potential satisfies
the Monge-Amp\`ere equation:
\begin{eqnarray}
K_{z_1\bar z_1}K_{z_2\bar z_2}-K_{z_1\bar z_2}K_{z_2\bar z_1}=\pm 1,
\end{eqnarray}
where the $+$ and $-$ sign stands for a $(4,0)$ and $(2,2)$ signature of
space-time resp. This can be integrated to the Plebanski action
\begin{eqnarray}
{\cal S}_{eff}=\int d^2 z_1 d^2 z_2 \, K\left\{\frac 1 3 \left(
K_{z_1\bar z_1}K_{z_2\bar z_2}-K_{z_1\bar z_2}K_{z_2\bar z_1}
\right)\mp 1 \right\}.
\end{eqnarray}

In \cite{ch,kiritsis,elias}, backgrounds consisting of one chiral and
one twisted chiral field were studied. UV finiteness is equivalent to
the Laplace equation
\begin{eqnarray}
K_{z\bar z}\pm K_{w\bar w}=0,
\end{eqnarray}
where again the $+$ and $-$ signs correspond to $(4,0)$ or $(2,2)$
signature resp. In this case the $N=2$ string describes a free scalar
field, with effective action:
\begin{eqnarray}
{\cal S}_{eff}=\frac 1 2 \int d^2 z d^2 w \left\{ K_z K_{\bar z}\pm K_w
K_{\bar w} \right\}.
\end{eqnarray}

Finally, having a background consisting of a semi-chiral multiplet
yields eq. (\ref{d4con}) as condition on the potential. Only when
$|F(r)|^2 |G(s)|^2=a$, with $a$ a positive real constant, can this be
integrated to an action cubic in the fields,
\begin{eqnarray}
{\cal S}_{eff}&=&\frac 1 3 \int d^2r d^2 s\, K\left\{K_{r\bar s}K_{r\bar s}\pm
a K_{rs}K_{\bar r\bar s}-\right. \nonumber\\
&&\left. \left(1\pm a \right)
K_{r\bar r}K_{s\bar s}\right\}.
\end{eqnarray}
For the former two cases, one can show that the holonomy group is contained
in $SU(2)$, hence these models describe gravitational instantons. Whether
or not this is also true for the latter case is being examined.

An important point which still has to be elucidated is the role of the
dilaton. In \cite{DD}, this was investigated for backgrounds consisting of
chiral and twisted chiral superfields. It was found that the geometry of
the $(2,2)$ super worldsheet implies the existence of four types
((anti-)chiral and twisted (anti-)chiral) of worldsheet curvatures, which
couple to fields satisfying the same constraints. A generalization of this
in the presence of semi-chiral superfields is presently under study.

\vspace{5mm}

\noindent {\bf Acknowledgments}:
We thank Chris Hull and Olaf Lechtenfeld for useful dicussions.
M.T.G. is suported in part by the NSF Grant No. PHY-92-22318.
M.M., A.S. and J.T. are supported in part by the European
Commission TMR programme ERBFMRX-CT96-0045 in which these authors are
associated
to K.\ U.\ Leuven.

\end{document}